\documentclass[12pt]{article}
\usepackage{times}
\usepackage{geometry}
\usepackage{makecell}
\usepackage[utf8]{inputenc}
\usepackage{enumitem,amssymb, wrapfig, xcolor}
\usepackage{ragged2e}\usepackage{graphicx}
\newlist{thematic}{itemize}{8}
\setlist[thematic]{label=$\square$}
\usepackage{pifont}

\usepackage{fourier} 
\usepackage{array}
\usepackage{makecell}

\newcommand\aaps{\ref@jnl{A\&AS}}%

\usepackage[utf8]{inputenc}
\usepackage[english]{babel}
\usepackage[numbers]{natbib}
\usepackage{graphicx}
\usepackage[firstpage]{draftwatermark}
\SetWatermarkLightness{0.9}
\SetWatermarkText{\includegraphics[angle=-45,width=2.0\textwidth]{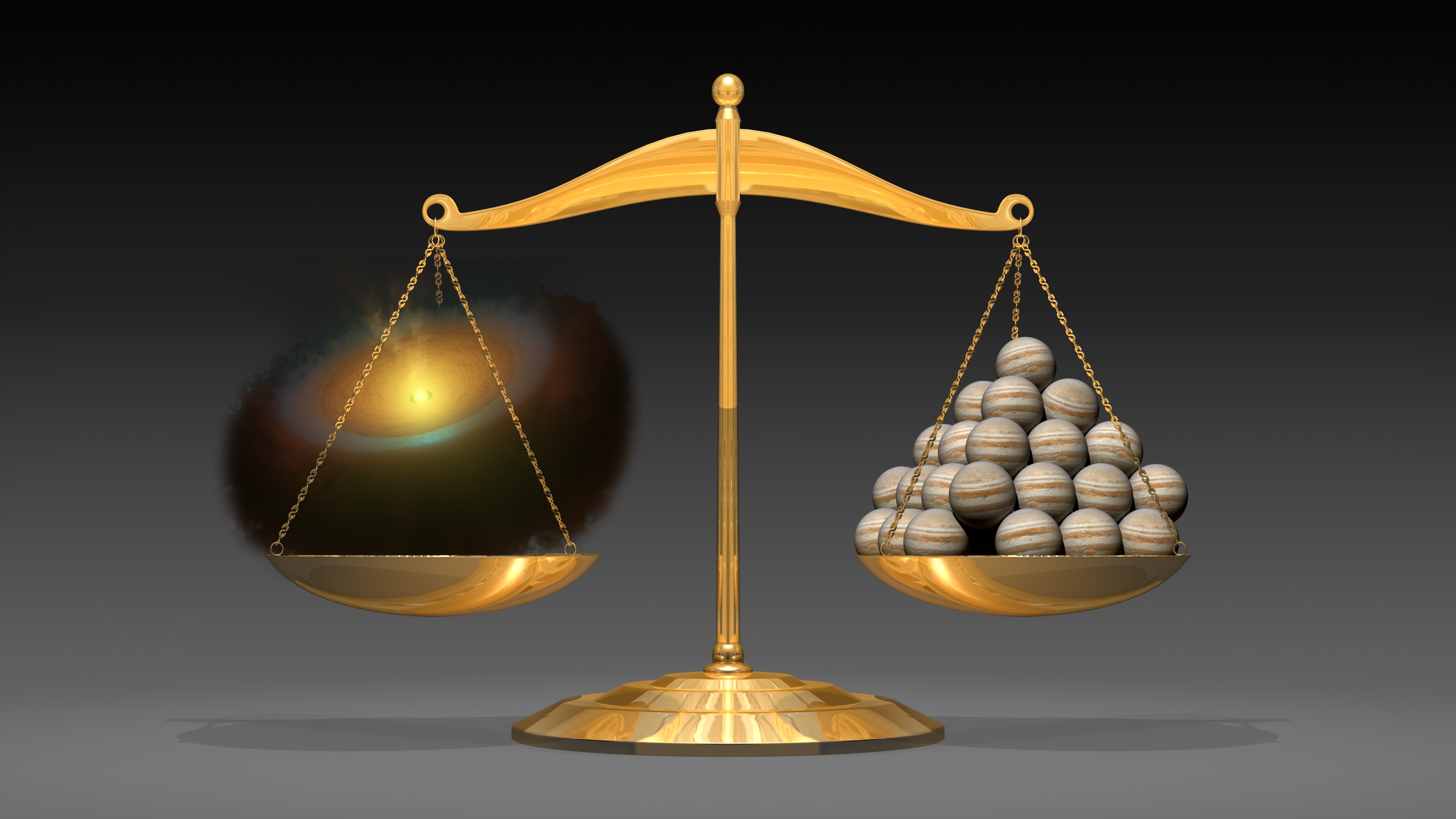}}
\usepackage{tcolorbox}
\definecolor{azure}{RGB}{240, 255, 255}

\begin{document}


\thispagestyle{empty}
\raggedright
\huge
\hspace{1cm}Astro2020 Science White Paper
\linebreak

The Disk Gas Mass and the Far-IR Revolution \linebreak
\normalsize

\noindent
\textcolor{white}{\textbf{Thematic Areas:} \hspace*{60pt} $\square$ Planetary Systems \hspace*{10pt} $\square$ Star and Planet Formation}

\textcolor{white}{\textbf{Principal Author:}}

\textcolor{white}{Name: Edwin A. Bergin}
 \linebreak						
\textcolor{white}{Institution:  University of Michigan}
 \linebreak
\textcolor{white}{Email: ebergin@umich.edu}
 \linebreak
\textcolor{white}{Phone:  734-615-8720}
 \linebreak
 
\textcolor{white}{\textbf{Co-authors:} }
  \linebreak
\textcolor{white}{Klaus M. Pontoppidan (STScI), Charles M. Bradford (Caltech-JPL), L. Ilsedore Cleeves (Univ. of Virginia), Neal J. Evans (Univ. of Texas), Maryvonne Gerin (Paris Obs.), Paul F. Goldsmith (Caltech-JPL), Quentin Kral (Paris Obs.), Gary J. Melnick (Harvard-Smithsonian CfA), Melissa McClure (University of Amsterdam), Karin \"Oberg (Harvard Univ.), Thomas L. Roellig (NASA Ames), Edward Wright (UCLA), Richard Teague (Univ. of Michigan), Jonathan P. Williams (Univ. of Hawaii), Ke Zhang (Univ. of Michigan)}

\newpage
\setcounter{page}{1}
\begin{tcolorbox}[width=\textwidth,colback={azure},colbacktitle=yellow,coltitle=blue]    
{\bf Abstract:}
The gaseous mass of protoplanetary disks is a fundamental quantity in planet formation. The presence of gas is necessary to assemble planetesimals, it determines timescales of giant planet birth, and it is an unknown factor for a wide range of properties of planet formation, from chemical abundances (X/H) to the mass efficiency of planet formation. The gas mass obtained from traditional  tracers, such as dust thermal continuum and CO isotopologues, are now known to have significant (1 - 2 orders of magnitude) discrepancies.  Emission from the isotopologue of H$_2$, hydrogen deuteride (HD), offers an alternative measurement of the disk gas mass.

Of all of the regions of the spectrum, the far-infrared stands out in that orders of magnitude gains in sensitivity can be gleaned by cooling a large aperture telescope to 8 K.  Such a facility can open up a vast new area of the spectrum  to exploration.  One of the primary benefits of this far-infrared revolution would be the ability to survey hundreds of planet-forming disks in HD emission to derive their gaseous masses.  For the first time, we will have statistics on the gas mass as a function of evolution, tracing birth to dispersal as a function of stellar spectral type.  These measurements have broad implications for our understanding of the time scale during which gas is available to form giant planets, the dynamical evolution of the seeds of terrestrial worlds, and the resulting chemical composition of pre-planetary embryos carrying the elements needed for life.   Measurements of the ground-state line of HD requires a space-based observatory operating in the far-infrared at 112~$\mu$m. \\
\end{tcolorbox}

{\bf Central question: What is the gaseous disk mass and its evolution during planet formation?}

Planet-forming disks are mostly gaseous in composition; at birth the gas carries $\sim$100$\times$ more mass than the solid dust particles. 
The  disk mass is the most fundamental quantity determining whether planets can form, and on what time scale. Estimates of disk gaseous masses are complicated by the fact that the first excited state of H$_2$ sits at 512~K above the ground state and is essentially unemissive at the temperatures that characterize much of the gaseous mass (i.e., 10--30~K). To counter this difficulty, the thermal continuum emission of the dust grains, or rotational lines of CO, are often used as a proxy for mass under the assumption they can be calibrated to trace the total gas mass \citep[see discussion in,][]{Bergin17}.  However, sensitive observations have demonstrated that grains have undergone substantial growth; in fact it is clear that significant dust mass is missing \citep{Manara16}. Further, recent ALMA surveys of well over 100 disks have compared gas masses measured from dust to CO-derived masses \citep{Ansdell16, Miotello17, Long17}. They find a systemic result that CO-derived masses are $10-100\times$ below that estimated from dust (assuming gas mass/dust mass = 100). Given that the dust masses are also a lower limit on the total mass, the gas mass of disks remains unknown by orders of magnitude despite decades of work.  

This leads to a gas mass conundrum such that either the CO abundance in disks may be orders of magnitude lower than in molecular clouds due to chemical and/or physical effects \citep{Krijt16, Eistrup17, Schwarz17, Yu17} or the planet-forming gaseous mass is dispersed within a few Myr \citep{Ansdell16}.  
\emph{This problem is pervasive, yet the effect is  the dominant term in the overall disk physical/chemical evolution. Furthermore, there are broad implications as all chemical measurements need the H$_2$ mass to determine abundances and the presence/absence of gas drives giant planet migration and influences the orbits of protoplanetary bodies 
\citep{Kominami02, Nelson18, McNally19}.}  

{\bf Hydrogen Deuteride as a New Tracer of Hidden H$_2$}

\begin{figure}
    \centering
    \includegraphics[width=0.9\textwidth]{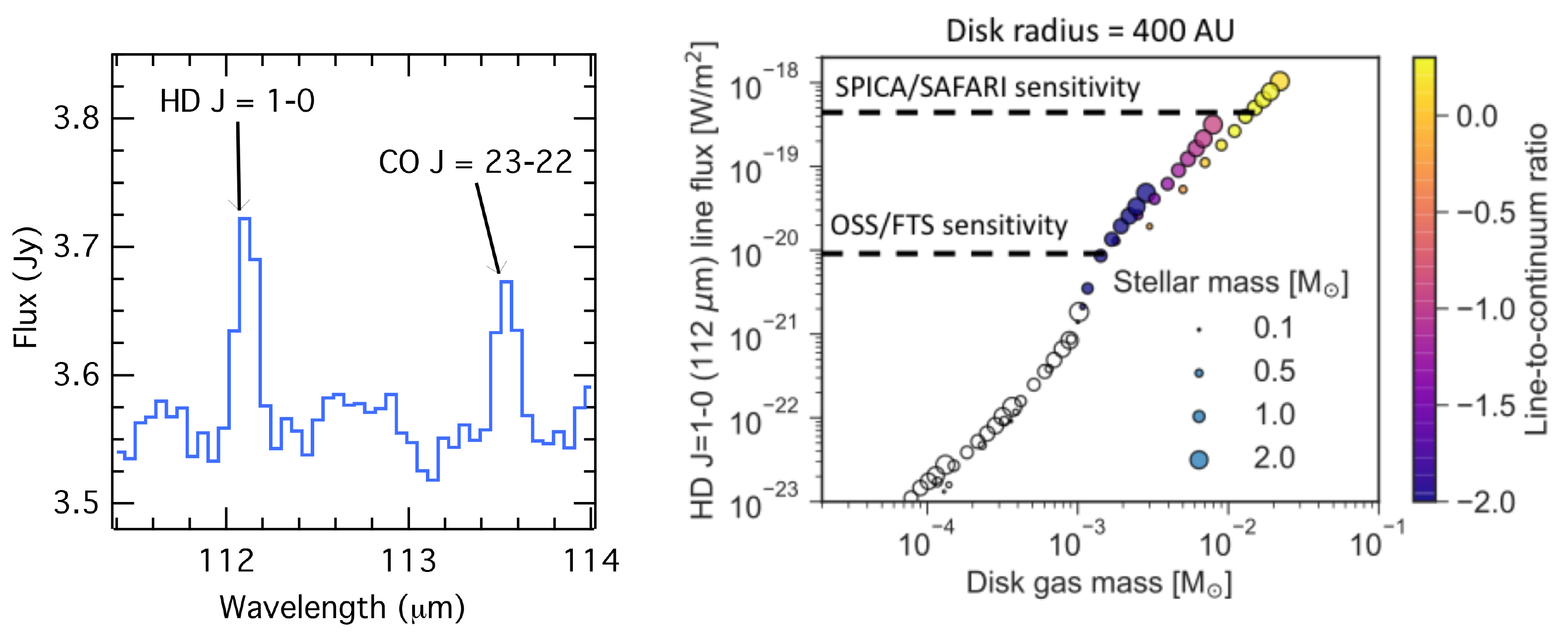}
    \caption{{\em Left:} \small HD emission at 112 $\mu$m towards TW Hya \citep{bergin_hd}. {\em Right:} Predicted emission as a function of disk gas mass and stellar mass for disk systems at the distance of Orion.  For reference, TW Hya is $\sim$7$\times$ closer than Orion. The logarithm of the line-to-continuum ratio (L/C) is also shown. Uncolored points have L/C $<$1\%. As the dust thermal continuum emission peaks near 100 $\mu$m, high spectral resolving powers, $R\gtrsim$ a few $10^{4}$, are needed to optimize the contrast enough for detection (L/C$\gtrsim$ a few \% for lower-mass disks, e.g. $M\lesssim 10^{-3} M_{\odot}$). Further, fully resolving the HD lines (in the absence of spatially resolved imaging), will help to decrease the error on the retrieved disk masses.}  
    \label{fig:1}
\end{figure}

The fundamental (J $=1 \rightarrow 0$) rotation transition of HD at 112~$\mu$m has been proposed as a much more robust measure of protoplanetary disk mass. While HD is a different molecular species, and therefore remains an indirect tracer of H$_2$, it may be as close as it is possible to get to a direct total mass tracer. Using Herschel, the gas mass of the TW Hya disk was measured by taking advantage of the fact that the lowest rotational transition of HD is $\sim 10^6$ times more emissive than the lowest transition of H$_2$ for a given gas mass at 20 K \citep{bergin_hd}. Due to Herschel’s limited lifetime, the only other deep HD observations obtained were toward six disks, with the result being two additional detections \citep{McClure16}. The TW Hya detection is shown in Fig.~\ref{fig:1} (left).

\begin{figure}
    \centering
    \includegraphics[width=0.95\textwidth]{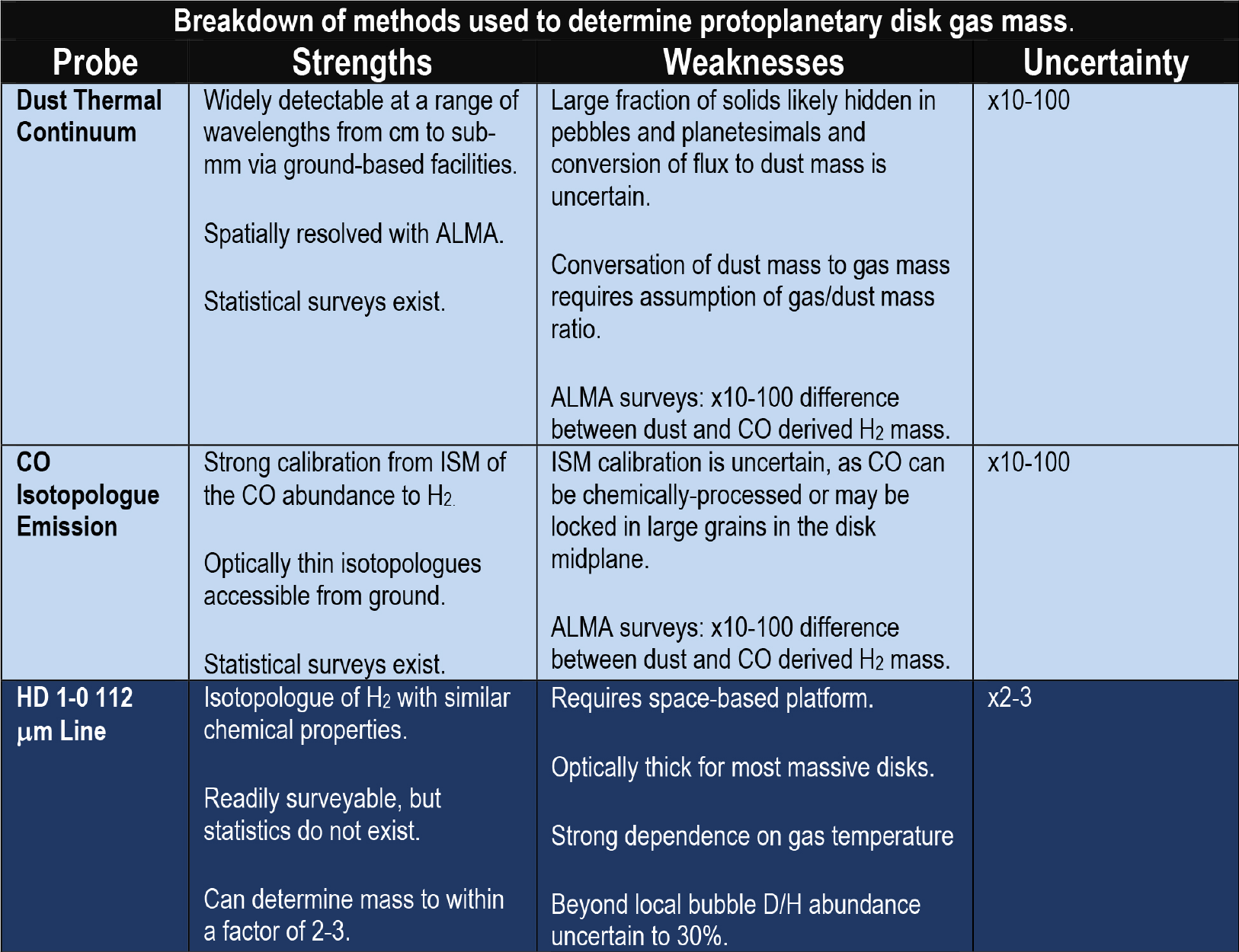}
    \label{fig:table}
\end{figure}
Table 1 provides a breakdown of the strengths and weaknesses of each indirect probe of the gas mass. Dust and CO-derived masses both have internal issues, but at present there is a factor of 10-100 discrepancy amongst these measurements. HD itself has two weaknesses. First, the atomic D/H ratio is used to set the calibration between the HD and H$_2$ mass. This varies by $\sim$30\% beyond the local bubble \citep{linsky98}.  This uncertainty is therefore much smaller than the order of magnitude uncertainties seen when comparing dust- or CO-estimated masses. Second, the HD gas mass has a strong dependence on the assumed gas temperature and is proportional to exp (-128/T$_{gas}$). This parameter dominates the uncertainty on HD as a mass tracer. However, ALMA observations of rare CO isotopologues can readily constrain models of  radial and vertical disk gas temperature in disk systems, independently of the CO abundance \citep{Schwarz16, Pinte18, Weaver18}. In addition, a far-IR telescope operating at 100 to 600~$\mu$m would provide access to CO isotopologues with  $J\ge 5$, tracing warmer gas. 

As seen in Fig.\ref{fig:1} (right),  models predict that the emission of HD directly measures mass (see also \citep{Trapman17}). With temperature constrained via other means, the disk gas mass can certainly be derived to within a factor of 2-3 \citep{bergin_hd,McClure16, Trapman17}. Given that the present day discrepancy encompasses orders of magnitude, this represents a major advance.

{\bf The Existing Landscape}

Detection of HD emission requires  being above the absorption of the Earth’s atmosphere.  The HIRMES instrument on SOFIA will have the ability to detect and partially resolve ($R\sim 100,000$) HD emission in the most massive and warmest disk systems with M$_{gas} \gtrsim 0.03$ M$_{\odot}$ (flux $\gtrsim 5\times 10^{-18}$ W/m$^2$). In addition, by spectrally resolving the line, one can use the fact that line width is dominated by temperature and Keplerian rotation to improve the retrieval of the total disk mass, given a disk model. That is, the line profile carries information on the mass distribution as a function of radius. Given the limited information we have at present, any detection by HIRMES will be invaluable and this instrument will pave the way for future studies. However, the SOFIA telescope is not cooled and thus does not capture the order of magnitude gains needed for a major step forward.

\begin{figure}
    \centering
    \includegraphics[width=0.7\textwidth]{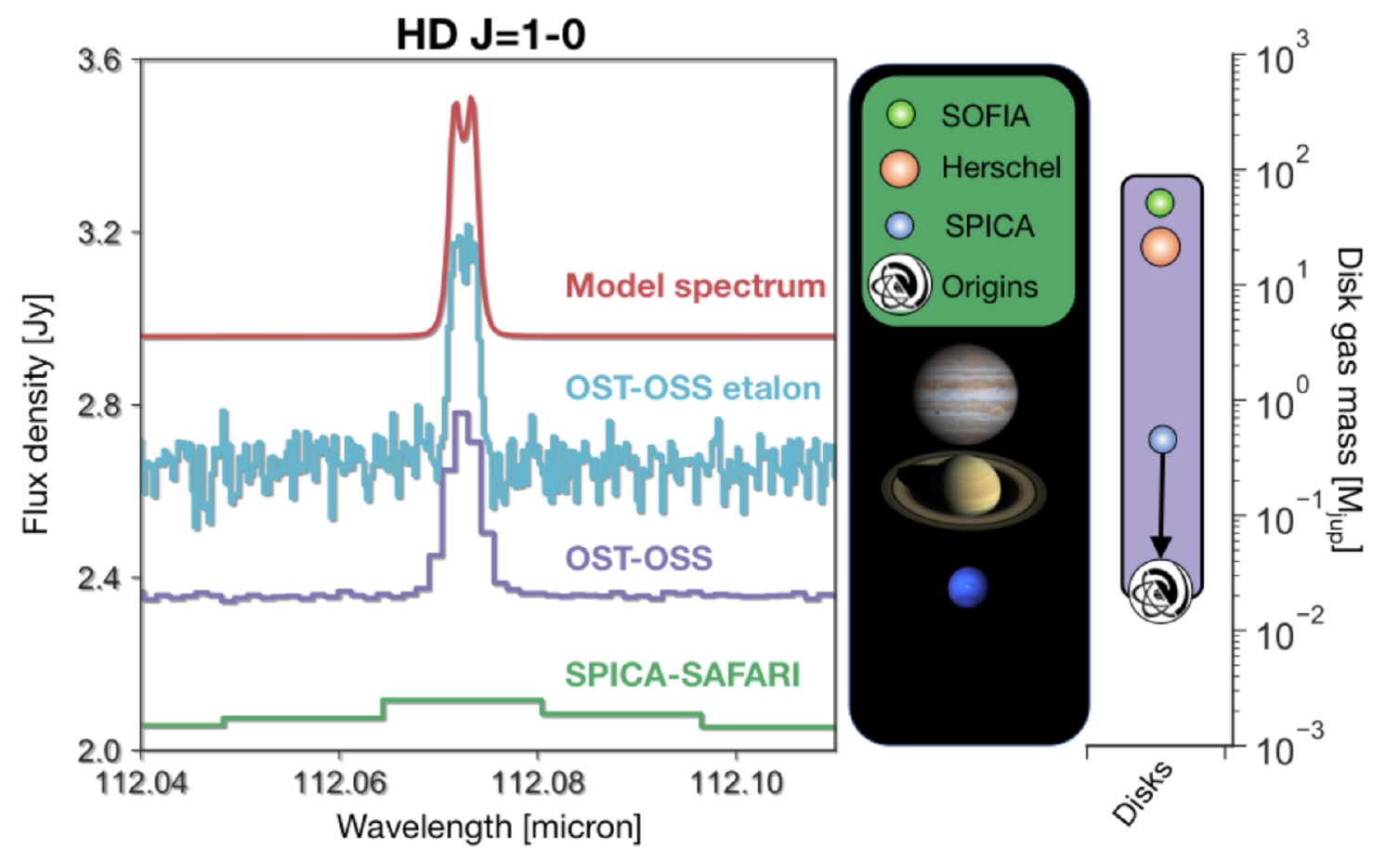}
    \caption{\small Model HD J $1 \rightarrow 0$ emission profile in a typical disk system along with spectra generated for three examples of spectral resolution (OSS-etalon - R $\sim 10^5$, OSS-(FTS) - R $\sim 10^4$, SPICA-SAFARI - R $\sim 3000$). The mass sensitivity is given in terms of Jupiter masses at 100 pc with example images of planets with similar masses show to the right.    These examples show the power of a cooled telescope (SOFIA vs Origins) and high spectral resolution (Origins vs SPICA) in terms of sensitivity to HD emission and gas mass.}
    \label{fig:2}
\end{figure}

In the field of disk studies we need to move beyond observations of single objects - we need to survey across all evolutionary stages and stellar mass. To obtain statistics on disks around stars of all masses, and to capture temporal disk evolution in a meaningful manner, we need to be sensitive to mass evolution. Thus any observatory  must be 
capable of detecting gas masses needed to make giant planets 
 ($< 10^{-3}$M$_{\odot}$) or $\sim 10^{-20}$ W/m$^{2}$ (5$\sigma$).  This also enables the detection of more massive systems at much larger distances to build a statistical sample.
 This requires a cooled space telescope. At present there are two options that have been explored in detail and provide a glimpse of the capabilities that are needed. SPICA, which is a 2.5m cooled aperture equipped with a spectrometer with intermediate spectral resolving power ($R \sim 3000$) and the Origins Space Telescope with a $\sim$5.9m cooled aperture and $R \sim 43,000$ at 112\,$\mu$m. SPICA is one of the three missions in competition for ESA's M5 opportunity and the Origins Space Telescope is a concept as part of NASA's decadal preparations.  Of these two the higher spectral resolving power (and larger aperture) of the Origins Space Telescope allows for HD line detections down to lower mass limits; the higher resolving power is particularly important when the disk continuum is bright, as is often the case.   These issues are  summarized in graphic form in Fig.\ref{fig:2}.

{\bf Recommendation}

The disk gas mass is a fundamental parameter for planet formation. It sets the timescales for giant planet formation and controls the kinematics of the system, including (proto-)planets.  During early stages, the presence of gas leads to the implantation of elements of life (Carbon, Oxygen, Hydrogen, Nitrogen, Sulfur) held in volatile molecules (e.g. CO, H$_2$O, ...) into the growing solids. Knowledge of the gas mass enables the determination of chemical abundances needed to track the incorporation of volatile elements in forming rocky and gas/giant planets, as a mandatory factor for habitability (for terrestrial worlds). Over the coming decade, knowledge of the planet-forming gas mass in the galaxy will slowly advance but a leap in our knowledge requires HD measurements, and this goal requires dedicated work toward sensitive spectroscopic observations at 112\,$\mu$m.   

We therefore recommend: 

\noindent  {\em the development of a large aperture far-infrared cooled space-borne observatory. The observatory must be sensitive to HD emission at 112\,$\mu$m with integrated fluxes of  $\;10^{-20}$~W~m$^{-2}$ to reach the lowest mass disks in the nearest star-forming regions and  to reach the several thousand disks in Orion. The spectral resolving power should be at least a few $\times$ 10$^{4}$ to have enough line-to-continuum contrast to detect lower-mass disks. This will enable detailed surveys of HD emission to obtain statistical information on this central quantity for the first time. 
}

\newpage
\bibliography{overleaf.bib}

\begin{thebibliography}{10}

\bibitem{Bergin17}
E.~A. {Bergin} and J.~P. {Williams}.
\newblock {The Determination of Protoplanetary Disk Masses}.
\newblock {\em Formation, Evolution, and Dynamics of Young Solar Systems},
  445:in press, 2017.

\bibitem{Manara16}
C.~F. {Manara}, G.~{Rosotti}, L.~{Testi}, A.~{Natta}, J.~M. {Alcal{\'a}}, J.~P.
  {Williams}, M.~{Ansdell}, A.~{Miotello}, N.~{van der Marel}, M.~{Tazzari},
  J.~{Carpenter}, G.~{Guidi}, G.~S. {Mathews}, I.~{Oliveira}, T.~{Prusti}, and
  E.~F. {van Dishoeck}.
\newblock {Evidence for a correlation between mass accretion rates onto young
  stars and the mass of their protoplanetary disks}.
\newblock {\em \aap}, 591:L3, June 2016.

\bibitem{Ansdell16}
M.~{Ansdell}, J.~P. {Williams}, N.~{van der Marel}, J.~M. {Carpenter},
  G.~{Guidi}, M.~{Hogerheijde}, G.~S. {Mathews}, C.~F. {Manara}, A.~{Miotello},
  A.~{Natta}, I.~{Oliveira}, M.~{Tazzari}, L.~{Testi}, E.~F. {van Dishoeck},
  and S.~E. {van Terwisga}.
\newblock {ALMA Survey of Lupus Protoplanetary Disks. I. Dust and Gas Masses}.
\newblock {\em \apj}, 828:46, September 2016.

\bibitem{Miotello17}
A.~{Miotello}, E.~F. {van Dishoeck}, J.~P. {Williams}, M.~{Ansdell},
  G.~{Guidi}, M.~{Hogerheijde}, C.~F. {Manara}, M.~{Tazzari}, L.~{Testi},
  N.~{van der Marel}, and S.~{van Terwisga}.
\newblock {Lupus disks with faint CO isotopologues: low gas/dust or high carbon
  depletion?}
\newblock {\em \aap}, 599:A113, March 2017.

\bibitem{Long17}
F.~{Long}, G.~J. {Herczeg}, I.~{Pascucci}, E.~{Drabek-Maunder}, S.~{Mohanty},
  L.~{Testi}, D.~{Apai}, N.~{Hendler}, T.~{Henning}, C.~F. {Manara}, and G.~D.
  {Mulders}.
\newblock {An ALMA Survey of CO Isotopologue Emission from Protoplanetary Disks
  in Chamaeleon I}.
\newblock {\em \apj}, 844:99, August 2017.

\bibitem{Krijt16}
S.~{Krijt}, F.~J. {Ciesla}, and E.~A. {Bergin}.
\newblock {Tracing Water Vapor and Ice During Dust Growth}.
\newblock {\em \apj}, 833:285, December 2016.

\bibitem{Eistrup17}
C.~{Eistrup}, C.~{Walsh}, and E.~F {van Dishoeck}.
\newblock {Molecular abundances and C/O ratios in chemically evolving
  planet-forming disk midplanes}.
\newblock {\em A\&A}, page in press (arXiv 1709. 07863), September 2017.

\bibitem{Schwarz17}
K.~R. {Schwarz}, E.~A. {Bergin}, L.~I. {Cleeves}, K.~{Zhang}, K.~I.
  {{\"O}berg}, and D.~{Blake}, G.~A.and~{Anderson}.
\newblock {Unlocking CO Depletion in Protoplanetary Disks I. The Warm Molecular
  Layer}.
\newblock {\em \apj}, page submitted, June 2017.

\bibitem{Yu17}
M.~{Yu}, N.~J. {Evans}, II, S.~E. {Dodson-Robinson}, K.~{Willacy}, and N.~J.
  {Turner}.
\newblock {Disk Masses around Solar-mass Stars are Underestimated by CO
  Observations}.
\newblock {\em \apj}, 841:39, May 2017.

\bibitem{Kominami02}
J.~{Kominami} and S.~{Ida}.
\newblock {The Effect of Tidal Interaction with a Gas Disk on Formation of
  Terrestrial Planets}.
\newblock {\em \icarus}, 157:43--56, May 2002.

\bibitem{Nelson18}
R.~P. {Nelson}.
\newblock {\em {Planetary Migration in Protoplanetary Disks}}, page 139.
\newblock 2018.

\bibitem{McNally19}
C.~P. {McNally}, R.~P. {Nelson}, S.-J. {Paardekooper}, and
  P.~{Ben{\'{\i}}tez-Llambay}.
\newblock {Migrating super-Earths in low-viscosity discs: unveiling the roles
  of feedback, vortices, and laminar accretion flows}.
\newblock {\em \mnras}, 484:728--748, March 2019.

\bibitem{bergin_hd}
E.~A. {Bergin}, L.~I. {Cleeves}, U.~{Gorti}, K.~{Zhang}, G.~A. {Blake}, J.~D.
  {Green}, S.~M. {Andrews}, N.~J. {Evans}, II, T.~{Henning}, K.~{{\"O}berg},
  K.~{Pontoppidan}, C.~{Qi}, C.~{Salyk}, and E.~F. {van Dishoeck}.
\newblock {An old disk still capable of forming a planetary system}.
\newblock {\em \nat}, 493:644--646, January 2013.

\bibitem{McClure16}
M.~K. {McClure}, E.~A. {Bergin}, L.~I. {Cleeves}, E.~F. {van Dishoeck}, G.~A.
  {Blake}, N.~J. {Evans}, II, J.~D. {Green}, T.~{Henning}, K.~I. {{\"O}berg},
  K.~M. {Pontoppidan}, and C.~{Salyk}.
\newblock {Mass Measurements in Protoplanetary Disks from Hydrogen Deuteride}.
\newblock {\em \apj}, 831:167, November 2016.

\bibitem{linsky98}
J.~L. {Linsky}.
\newblock {Deuterium Abundance in the Local ISM and Possible Spatial
  Variations}.
\newblock {\em \ssr}, 84:285--296, April 1998.

\bibitem{Schwarz16}
K.~R. {Schwarz}, E.~A. {Bergin}, L.~I. {Cleeves}, G.~A. {Blake}, K.~{Zhang},
  K.~I. {{\"O}berg}, E.~F. {van Dishoeck}, and C.~{Qi}.
\newblock {The Radial Distribution of H$_{2}$ and CO in TW Hya as Revealed by
  Resolved ALMA Observations of CO Isotopologues}.
\newblock {\em \apj}, 823:91, June 2016.

\bibitem{Pinte18}
C.~{Pinte}, F.~{M{\'e}nard}, G.~{Duch{\^e}ne}, T.~{Hill}, W.~R.~F. {Dent},
  P.~{Woitke}, S.~{Maret}, G.~{van der Plas}, A.~{Hales}, I.~{Kamp}, W.~F.
  {Thi}, I.~{de Gregorio-Monsalvo}, C.~{Rab}, S.~P. {Quanz}, H.~{Avenhaus},
  A.~{Carmona}, and S.~{Casassus}.
\newblock {Direct mapping of the temperature and velocity gradients in discs.
  Imaging the vertical CO snow line around IM Lupi}.
\newblock {\em \aap}, 609:A47, January 2018.

\bibitem{Weaver18}
E.~{Weaver}, A.~{Isella}, and Y.~{Boehler}.
\newblock {Empirical Temperature Measurement in Protoplanetary Disks}.
\newblock {\em \apj}, 853:113, February 2018.

\bibitem{Trapman17}
L.~{Trapman}, A.~{Miotello}, M.~{Kama}, E.~F. {van Dishoeck}, and
  S.~{Bruderer}.
\newblock {Far-infrared HD emission as a measure of protoplanetary disk mass}.
\newblock {\em \aap}, 605:A69, September 2017.

\end{thebibliography}

\end{document}